\documentclass[twocolumn]{aastex631}

\usepackage{lipsum}

\received{March 1, 2021}
\revised{April 1, 2021}
\accepted{\today}

\submitjournal{ApJ}

\shorttitle{Hi-C 2.1 Nanojets}
\shortauthors{Patel \& Pant}
\graphicspath{{./}{figures/}}
\begin{document}

\title{Hi-C 2.1 Observations of Reconnection Nanojets}

\correspondingauthor{Vaibhav Pant}
\email{vaibhav.pant@aries.res.in}

\author[0000-0001-8504-2725]{Ritesh Patel}
\affiliation{Indian Institute of Astrophysics, 2nd Block, Koramangala, Bangalore, 560034, India}
\affiliation{Aryabhatta Research Institute of Observational Sciences, Nainital, 263001, India }

\author[0000-0002-6954-2276]{Vaibhav Pant}
\affiliation{Aryabhatta Research Institute of Observational Sciences, Nainital, 263001, India }

\begin{abstract}

One of the possible mechanisms for heating the solar atmosphere is the magnetic reconnection occurring at different spatio-temporal scales. The discovery of fast bursty nanojets due to reconnection in the coronal loops has been linked to nanoflares and considered as possible mechanism for coronal heating. The occurrence of these jets mostly in the direction inwards to the loop were observed in the past. In this study, we report ten reconnection nanojets, four with directions inward while six moving outward to the loop, in observations from High-resolution Coronal Imager 2.1 (Hi-C 2.1) and Atmospheric Imaging Assembly (AIA) onboard Solar Dynamics Observatory (SDO). {We determined the maximum length, spire width, speed, and lifetimes of these jets and studied their correlations. We found that outward jets with higher speeds are longer in length and duration while the inward jets show opposite behaviour. Average duration of the outward jets is $\approx$42 s and inwards jets is $\approx$24s. We identified jets with subsonic speeds below 100 km s$^{-1}$ to high-speed over 150 km s$^{-1}$.}
These jets can be identified in multiple passbands of AIA extending from upper transition region to the corona suggesting their multi-thermal nature. 
% A differential emission measure (DEM) analysis yielded that the coronal plasma temperature of the jets have been estimated to an average logT$\sim$6.3. To the best of our knowledge, this is the first observational study on the outward moving reconnection nanojets in the solar corona.

\end{abstract}

\keywords{Solar corona (1483) --- Solar coronal transients (312) --- Solar magnetic reconnection (1504)}

\section{Introduction} 
    \label{sec:intro}

The flares have been observed in the solar atmosphere caused by magnetic reconnection at different spatial scales \citep{Guidoni2016ApJ, Benz2017LRSP, Patel2020A&A}. These reconnections occurring at various spatio-temporal scales are also considered as one of the mechanisms responsible for the coronal heating. {These reconnections contribute to coronal heating where flare energies follow a power-law distribution, $E \sim W^{-1.8}$, where W is the total flare energy \citep{Hudson1991SoPh}.} The continuous footpoint motion of the magnetic field lines leads to the braiding of magnetic field lines, that cause magnetic reconnection and formation of small scale current sheets eventually leading to nanoflares \citep{Parker1988ApJ, Ballegooijen2011ApJ} or recurrent plasma outflows \citep{Pant2015ApJ}. According to the Parker model, the small misalignment in the braided magnetic field lines within a { flux tube} cause magnetic reconnection. The stored magnetic energy is released and converted to the thermal, and kinetic energies which are responsible for acceleration of particles. 

Magnetic reconnections often lead to the formation of jets observed as inverted Y-shaped structures first observed in solar corona in X-ray wavelengths \citep[eg:][]{Shimojo1996PASJ, Shimojo2000ApJ, Savcheva2007PASJ}. Coronal jets are also observed in extreme ultraviolet (EUV) wavelengths \citep[eg:][]{Nistico2010, Chandrashekhar2014A&A, Liu2016ApJ, Sarkar2016SoPh}. These jets tend to show average projected lengths and widths in the range of 10-400 Mm and 5-100 Mm respectively with speeds of 10-1000 km s$^{-1}$ and average lifetime in the range of few tens of minutes \citep{Shimojo2000ApJ, Mulay2016A&A}. In addition to these, the lower range of these jets, termed as jetlets, are know to have sizes about three times smaller than average coronal jets \citep{Raouafi2014ApJ, Panesar2018ApJ}. Lower in the atmosphere, the jets are observed in the transition region with relatively smaller spatio-temporal scales \citep{Tian2014Sci, Narang2016SoPh, Samanta_2017, Chen2019ApJ}.
The jets observed at chromospheric and transition region heights have projected lengths and widths of 1-11 Mm, 100-400 km respectively with speeds in the range of 5-250 km s$^{-1}$ and 20-500 s lifetime \citep{DePontieu2011Sci, Tian2014Sci, Samanta2019Sci}. Using the high resolution Extreme Ultraviolet Imager {\citep[EUI:][]{EUI2020A&A}} onboard the recently launched Solar Orbiter, \citet{HouMicrojet2021ApJ} reported the presence of even small-scale microjets in the quiet region of the solar atmosphere. These microjects were found to have projected speed, width, maximum length, and average lifetime, of  62 km s$^{-1}$, 1.0 Mm, 7.7 Mm, and 4.6 minutes, respectively.

Recent observation of jet-like structures called nanojets observed in the reconnecting, curved-braided magnetic field lines in the solar atmosphere, provided a signature of nanoflare based heating \citep{Antolin2021NatAs}. These very fast-speed (100-200 km s$^{-1}$) and shot-lived ($<$15 s) jets are much smaller in size as compared to long-lived jet-like events observed earlier \citep[][and references therein]{Panesar2019ApJ, Raouafi2016SSRv} with origins from photospheric magnetic flux cancellation. The earlier studies based on flux cancellation were mostly based on the observations from Atmospheric Imaging Assembly \citep[AIA:][]{aia}, and Helioseismic and Magnetic Imager \citep[HMI:][]{HMI2012} of Solar Dynamics Observatory (SDO) and Interface Region Imaging Spectrograph \citep[IRIS:][]{IRIS2014SoPh}. Such flux cancellation jets were also observed using High-resolution Coronal imager \citep[Hi-C 2.1;][]{HiC2019SoPh} by \citet{Panesar2019ApJ}. {The high resolution imaging from Hi-C 2.1 also enabled the estimation of widths of fine strands of coronal loops \citep{Williams2020aApJ, Williams2020bApJ}, analysis of small-scale explosive events by flux cancellation \citep{Tiwari2019ApJ}, and identification of miniature filament eruption \citep{Sterling2020ApJ}.}

An interesting phenomena observed by \citet{Antolin2021NatAs} was the presence of jets directed inwards to curvature of the coronal loop. To explain the asymmetric nature of nanojets, \citet{Pagano2021A&A} performed analytical and numerical magnetohydrodynamic (MHD) simulations. Their analysis explained that the inward moving jets are more frequent and energetic than those moving outwards (away from the curvature of coronal loops). {In a recent study it was found that the Kelvin-Helmholtz and Rayleigh-Taylor instabilities are the drivers for the reconnection in coronal loops leading to nanojets \citep{2022ApJ...934..190S}.} In this Paper, we present the observational study of outward and inward nanojets using Hi-C data complemented by the multi-wavelength observations from the AIA.
This paper is organised as follows: In Section \ref{sec:obs} we describe the datasets used for this study. The analysis and results are presented in Section \ref{sec:results} followed by conclusions and a discussion in Section \ref{sec:conclusion}.

\section{Observations} 
    \label{sec:obs}

The second flight of Hi-C launched on May 29, 2018 observed an active region (AR; AR 12712) for $\approx$5 minutes from 18:56:26-19:01:56 UT in 172 \AA\ emission passband. The high spatial- (0.129" pixel$^{-1}$) and temporal-resolution (4.4 s) of Hi-C 2.1 \citep{HiC2019SoPh} provided an unprecedented view of this active region. We identified ten reconnection nanojets at different locations within the coronal loops in the Hi-C field of view (FOV), having direction inwards as well as outwards. 

The same were also identified in relatively low resolution in AIA 171 and 304 \AA\ channels with signatures in 193 and 131 \AA\ for a few. We have also used 45 s HMI line-of-sight (LOS) magnetograms of the FOV to negate the presence of any possible photospheric flux cancellations. The coordinated IRIS observations were not available for this part of the FOV.

\begin{figure*}[!ht]
    \centering
    \includegraphics[width=0.9\textwidth,clip=]{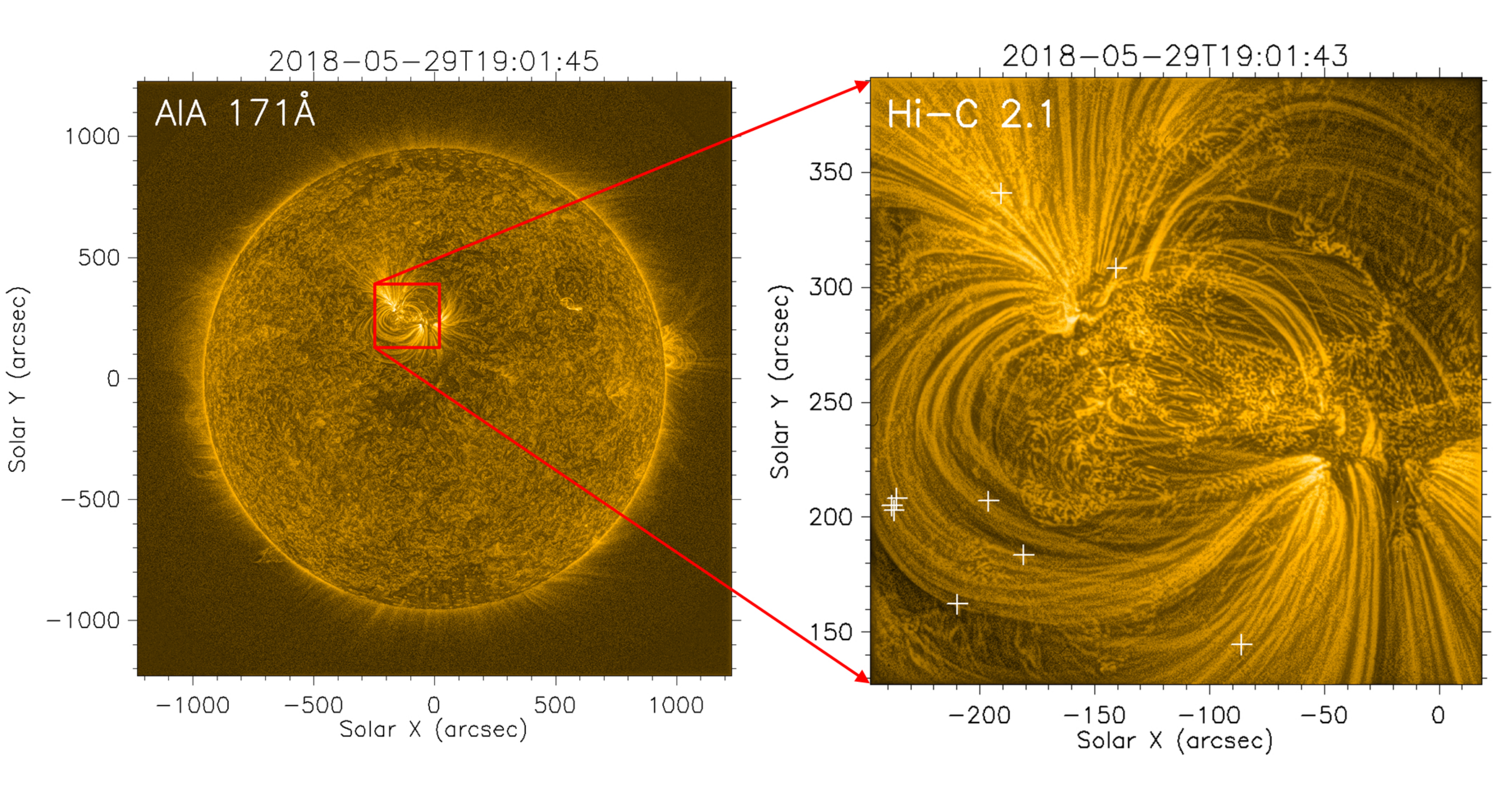}
    
     \centerline{    
      \hspace{0.3\textwidth}  \color{black}{(a)}
      \hspace{0.38\textwidth}  \color{black}{(b)}
    %   \hspace{0.29\textwidth}  \color{black}{(c)}
         \hfill}
         
    % \includegraphics[width=0.8\textwidth,clip=]{figures/braid_ridge.pdf}
    
    % \centerline{    
    %   \hspace{0.4\textwidth}  \color{black}{(d)}
    %   \hspace{0.3\textwidth}  \color{black}{(e)}
    %      \hfill}
         
    \caption{(a) Full-disk image of the Sun at 171 \AA\ of AIA. Red box represents the Hi-C FOV that is shown in the (b). The locations of the identified nanojets are maked by `+' symbol in the Hi-C FOV in white.} 
    % (c) The region of interest (ROI) of the identified nanojets is highlighted in cyan dashed box with zoomed view. S1 and S2 are the two slices for which the time evolution is shown in (d) and (e) images of bottom pannel respectively. The starting time of the distance-time maps are shown as T$_0$ at the top of each plot. \\ (Animation of this Figure is available.)}
    \label{fig:composite}
\end{figure*}
    
\section{Analysis and Results} 
    \label{sec:results}

\subsection{Overview}
The top panel of Figure \ref{fig:composite} shows AIA 171 \AA\ image with Hi-C FOV outlined with a red box. The Hi-C image processed with multi-scale Gaussian normalisation \citep[MGN;][]{MGN2014SoPh} shown in panel (b) of the same Figure. {The locations of all the identified nanojets have been marked with white `+' symbol in Figure \ref{fig:composite}b. A total of ten jets were identified on the edges of the coronal loops out of which six were directed outwards while four were directed inwards. To substantiate that these jets do not have origin due to the photospheric magnetic flux cancellation, we used HMI observations. Average magnetic field was estimated at the location near the base of the jets and the time evolution was studied. It was found that unlike the previous studies, these observed jets do not originate from the photospheric flux cancellation. Moreover, the observed magnetic flux remains near to the noise level of the HMI instrument \citep{Pesnell2012SoPh}.

\begin{deluxetable*}{ccllrrrrl}
\tablenum{1}
\tablecaption{Physical properties of the observed nanojets \label{tab:jet_prop}}
\tablewidth{0pt}
\tablecolumns{9}
\tablehead{
\colhead{Nanojet ID} & \colhead{Start Time (UT)} & \colhead{x\arcsec} & \colhead{y\arcsec} & \colhead{Speed (km s$^{-1}$)} & \colhead{Spire Length (km)} & \colhead{Spire Width (km)} & \colhead{Lifetime (s)} & \colhead{Direction}}
% \decimalcolnumbers
% \decimals
\startdata
N1 & 19:00:46 & -209.6 & 161.7 & 152 $\pm$ 24 & 4508 $\pm$ 231 & 1180 $\pm$ 118 & 40 & {\bf Outward} \\
N2 & 19:01:30 & -209.6 & 161.7 & 234 $\pm$ 29 & 4100 $\pm$ 179 & 1440 $\pm$ 157 & 40 & {\bf Outward}\\
N3 & 18:57:21 & -86.02 & 144.1 & 203 $\pm$ 31 & 3255 $\pm$ 201 & 1365 $\pm$ 77 & 79 & {\bf Outward}\\
N4 & 18:58:58 & -235.9 & 207.7 & 35 $\pm$ 9 & 2433 $\pm$ 352 & 1231 $\pm$ 196 & 31 & {\bf Outward} \\
N5 & 18:59:02 & -190.5 & 340.5 & 63 $\pm$ 14 & 2555 $\pm$ 155 & 1245 $\pm$ 236 & 26 & {\bf Outward} \\
N6 & 19:00:29 & -140.4 & 307.9 & 56 $\pm$ 5 & 924 $\pm$ 141 & 766 $\pm$ 84 & 40 & {\bf Outward} \\
% \hline
N7 & 18:57:18 & -238.2 & 204.5 & 46 $\pm$ 14 & 1952 $\pm$ 284 & 1207 $\pm$ 207 & 31 & {\bf Inward} \\
N8 & 18:57:18 & -237.2 & 202.5 & 47 $\pm$ 12 & 1690 $\pm$ 215 & 1011 $\pm$ 130 & 13 & {\bf Inward} \\
N9 & 18:56:52 & -196.05 & 206.5 & 62 $\pm$ 5 & 1573 $\pm$ 143 & 1175 $\pm$ 131 & 9 & {\bf Inward}  \\
N10 & 18:59:33 & -180.8 & 183.2 & 46 $\pm$ 6 & 2822 $\pm$ 117 & 1114 $\pm$ 129 & 40 & {\bf Inward} \\
\hline
% average $\pm$ 1 $\sigma_{av}$ & & 39.5 $\pm$ 0.7 & 4304 $\pm$ 289 & 635 $\pm$ 120 & 193 $\pm$ 58 \\
\enddata
% \tablecomments{}
\end{deluxetable*}

\begin{figure*}[!ht]
    \centering
    \includegraphics[width=1.05\textwidth,clip=]{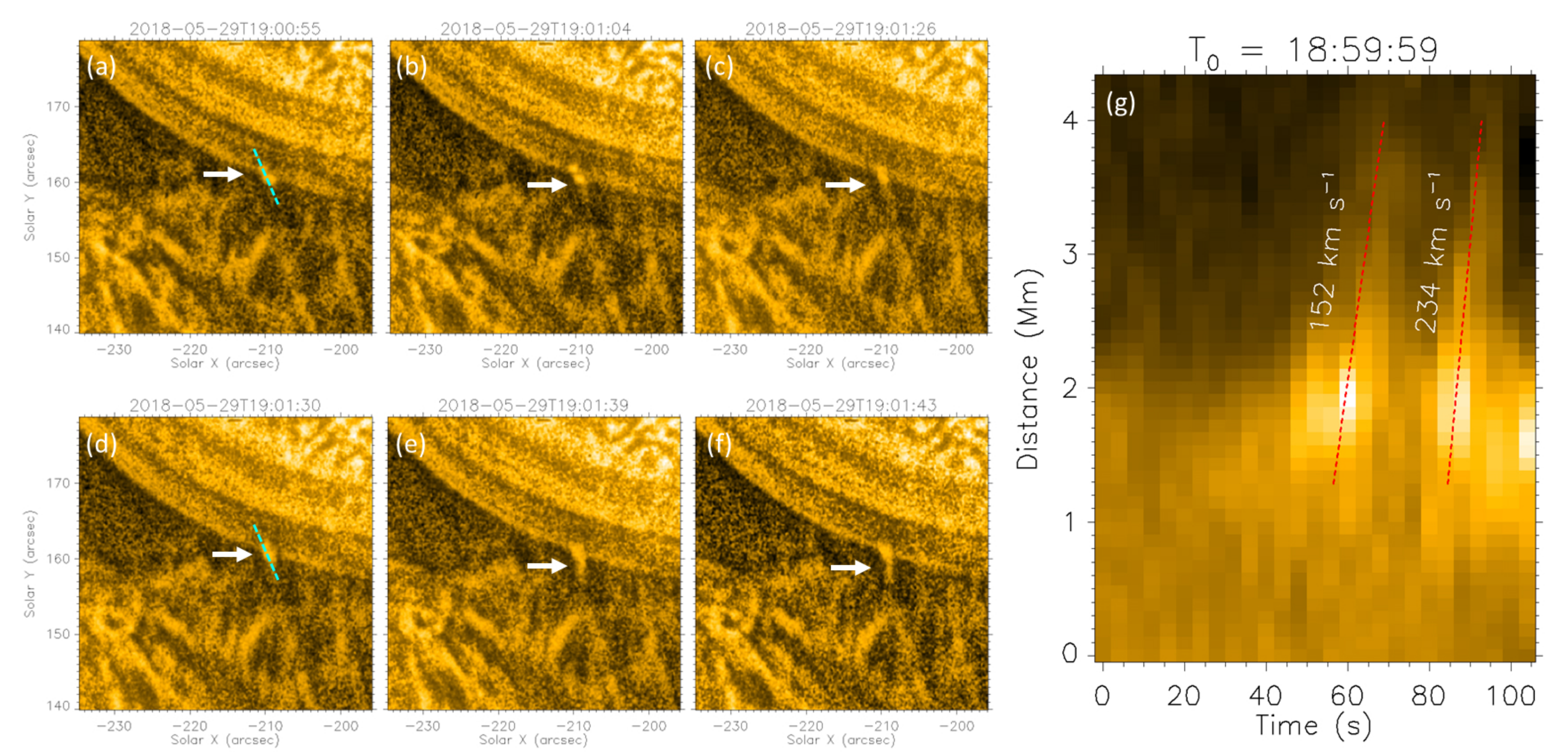}
    \caption{Time evolution of two outward moving nanojets, 1 and 2. (a) to (c) shows snapshots in Hi-C FOV for nanojet 1 while (d) to (f) shows snapshots for nanojet 2. The white arrow points to the location of the jets. The time-distance plot is generated along the dashed cyan line shown in (a) and (d). (g) The time-distance plot for the two nanojets with starting time shown as T$_0$ at the top of the plot. \\ ({Animation of this Figure is available showing the time evolution of jets along with panel (g).)}}
    \label{fig:reg2jets}
\end{figure*}

The properties of these nanojets projected in the plane-of-sky are tabulated in Table \ref{tab:jet_prop}. The spire length and width are measured as described in \citet{Panesar2019ApJ}. {The third and fourth columns correspond to the location of the base of the jets in arcseconds. Nanojets N1 to N6 are the outward moving jets while N7 to N10 are the inward moving jets.} The errors quoted in the Table are statistical in nature determined by repeated measurements of parameters. We found that the average speed of the outward jets is of an order of few hundreds km s$^{-1}$ with few having sub-sonic speeds. The inward jets on the other hand have average speed of $\approx$50 km~s$^{-1}$. We could identify outward jets with length ranging from 924 km to 4508 km which is more than twice the length of the inward jets. The average width of both the classes of jets is $\approx$1100 km. We could identify jets with lifetimes as small as 8 s to 79 s with an average duration of 35 s.
Out of the ten jets, we present the morphological and kinematical evolution of five nanojets:}

\subsection{Nanojets N1 and N2}
Figure \ref{fig:reg2jets} shows examples of jets moving outward from the coronal loop. An artificial slice is placed along the direction of propagation of the jet shown as cyan dashed line in panels (a) and (d). The snapshots corresponding to the two jets occurring from the same spot but differing in time are shown in panels (a) to (c) and (d) to (f) respectively.
% Figure \ref{fig:composite}c shows a nanojet observed moving outward from the coronal loop. An artificial slice S2 is placed along the direction of propagation of the nanojet. Another slice S1 is placed perpendicular to the loop strand just ahead of the nanojet to identify the strand movements. The time evolution corresponding to the two slices are shown in the bottom panel of the same Figure. It can be seen in the Figure \ref{fig:composite}d that two  strands appear to come closer with time. The path of the incoming strands are marked with dashed red line whereas the white-dashed line marks the time after which the two nanojets appeared. This suggests that the nanojets arise due to the reconnection of the incoming strands of the coronal loop.

\begin{figure*}
    \centering
    \includegraphics[width=1.02\textwidth,clip=]{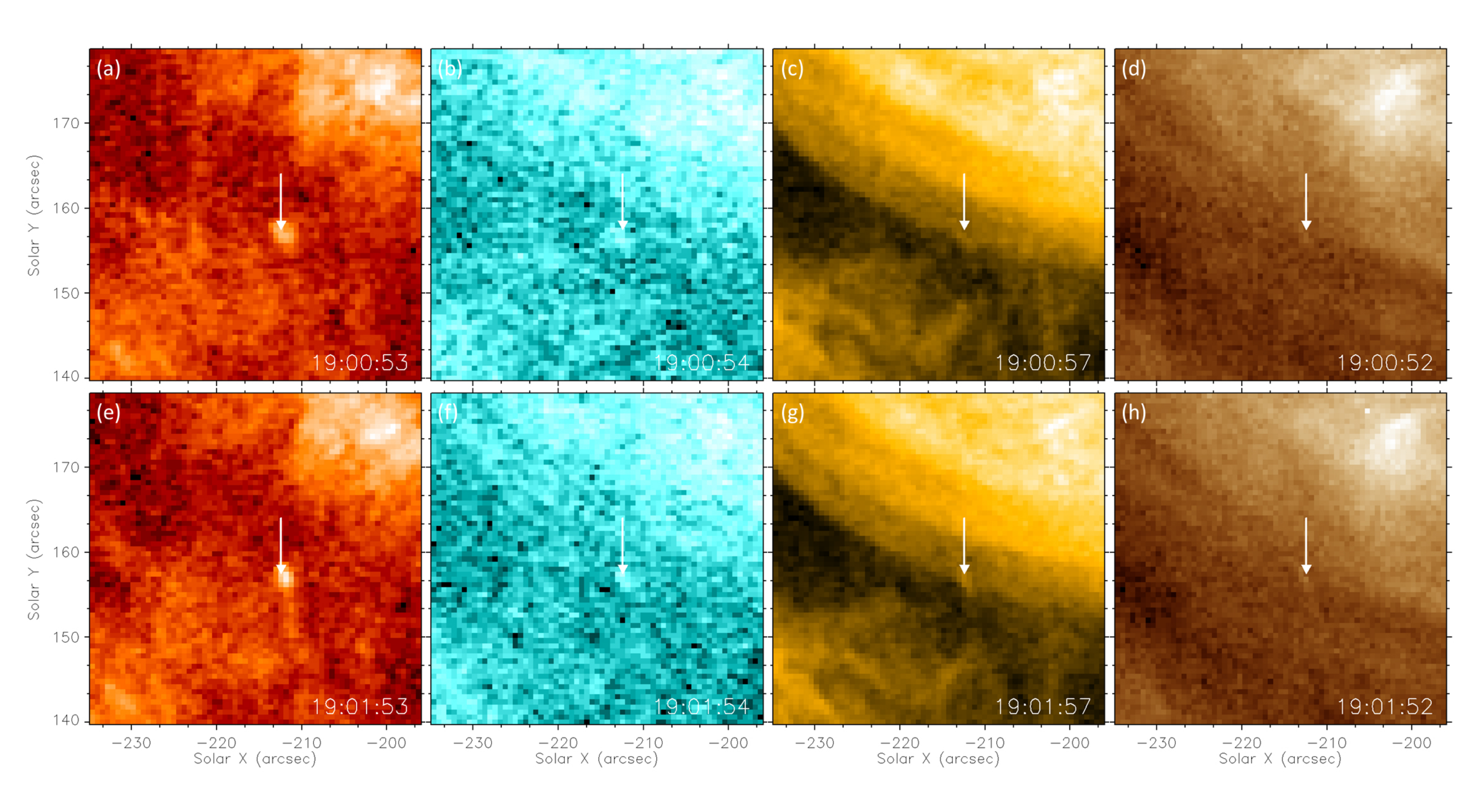}
    \caption{Multi-wavelength AIA observations of the two nanojets, N1 and N2, observed in Hi-C FOV. The {\it upper} and {\it lower} panels corresponds to the two nanojets observed in AIA 304, 131, 171 and 193 \AA\ respectively from left to right.}
    \label{fig:aia}
\end{figure*}

The time-distance map along the slice is shown in the  Figure \ref{fig:reg2jets}g where the two ridges mark the presence of two successive jets from the same location. {The linear fitting is done on the ridges in the distance-time map for multiple times (as many as ten times) to cover the width of the ridge and account for the manual fitting. The panel (g) corresponding to Figure \ref{fig:reg2jets} (and successive such plots) show the representative line which is an average of the multiple measurements.}
These two jets are one of the longest and fastest of the observed jets in this dataset. {The duration of jet N2 extends beyond the available Hi-C observation which covers 26 s. AIA 171~\AA\ image where it is last identified is used to quantify its complete lifetime. }

The multi-wavelength observations of the two jets, N1 and N2, are shown in the two respective panels of Figure \ref{fig:aia}. The images from left to right in the two panels correspond AIA 304 \AA\ (Figure~\ref{fig:aia}a,e) representing lower transition region temperature (logT $\approx$ 4.8-5), upper transition region temperatures (logT $\approx$5.7-5.9) probed by AIA 131 \AA\ (Figure~\ref{fig:aia}b,f) and 171 \AA\ (Figure \ref{fig:aia}c,g). The coronal temperature (logT $\approx$ 6.2-6.3) is shown in the panels d and h of the same Figure with AIA 193 \AA\ observations. In all the images the base of the observed jets is marked with white arrow. The observation of jets over a range of AIA passbands suggests that these are multi-thermal in nature reaching the coronal temperatures.

\subsection{Nanojets N4, N7 and N8}
\begin{figure*}
    \centering
    \includegraphics[width=1.02\textwidth,clip=]{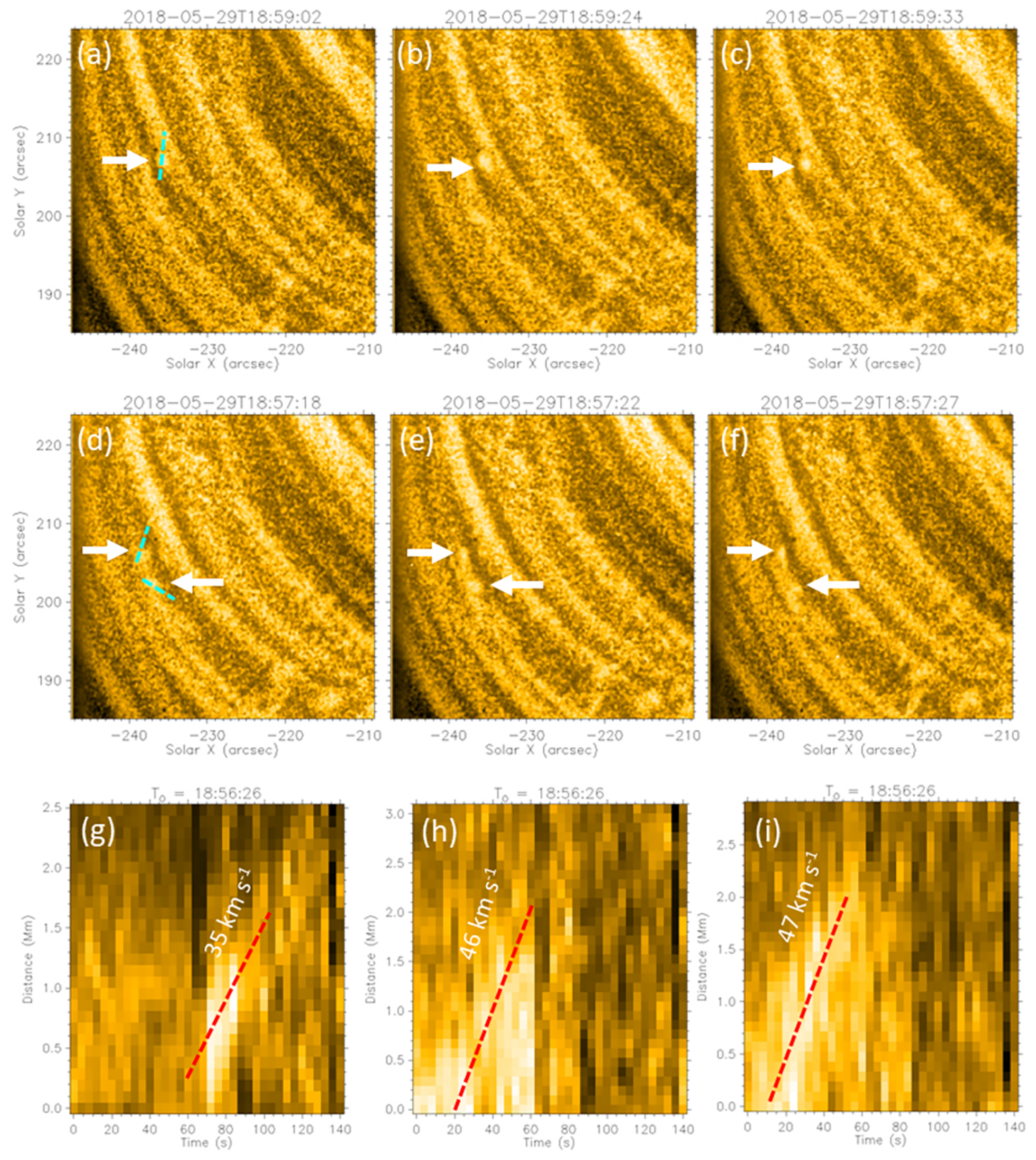}
    \caption{(a) to (c) shows time evolution snapshots of an outward nanojets, N4, in Hi-C FOV while (d) to (f) shows snapshots for nanojets N7 and N8 which move inwards. The white arrow points to the location of the jets. The time-distance plot is generated along the dashed cyan line shown in (a) and (d). (g) to (i) shows the time-distance plots for the jets N4, N7, and N8 respectively with starting time shown as T$_0$ at the top of the plots. \\ ({ Animation of this Figure is available.)}}
    \label{fig:reg5jets}
\end{figure*}

{Figure \ref{fig:reg5jets} shows snapshots of a region where three nanojets are observed. An outward moving jet, N4, is shown in panels (a) to (c) while N7 and N8 appear to originate from the same spot but in opposite directions. In these projected plane it appears both are directed inwards one moving upwards (N7) while the other (N8) downwards. The white arrow in all the panels points towards the location of these jets. The cyan dashed line shown in panels (a) and (d) are the slices along which the time-distance plots are made to study their temporal evolution. The time-distance plots corresponding to N4, N7 and N8 are shown in panels (g), (h) and (i) of the same figure. The bright ridge in these plots show the movement of the jet along the cyan slices shown in (a) and (d) panels. All the three jets are found to be having speeds less than 50 km s$^{-1}$.}

\begin{figure*}
    \centering
    \includegraphics[width=.95\textwidth,clip=]{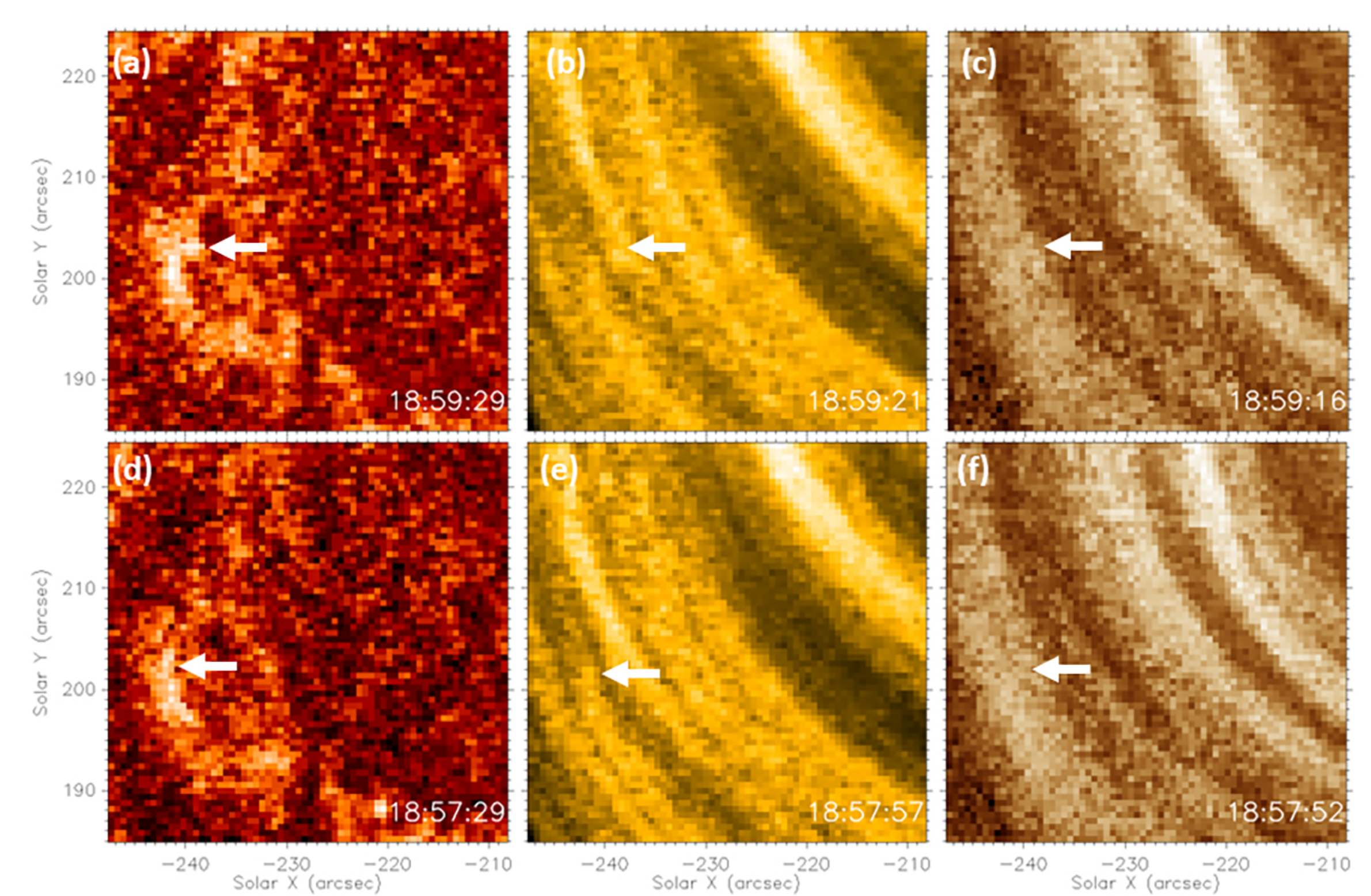}
    \caption{Multi-wavelength AIA observations of two nanojets, N4 and N7, observed in Hi-C FOV. The jet N8 being small could not be distinguished from the background in AIA resolution. The {\it upper} and {\it lower} panels corresponds to the two jets observed in AIA 304, 171 and 193 \AA\ respectively from left to right. (c) The jet N3 is barely visible in 193 \AA\ but no sign for jet N7 could be seen in (f) for the same channel.}
    \label{fig:reg5aia}
\end{figure*}

{Two of these jets, N4 and N7, showed some signatures in multiple passbands of AIA (Figure~\ref{fig:reg5aia}).  N4 could be seen as small brightening in 304 and 171 \AA\ passbands with a very week signature in the hotter 193 \AA\ channel of AIA (Figure \ref{fig:reg5aia}~a,b,c). On the other hand, N7 shows signatures in AIA 304 and 171 \AA\ channels while no signature is visible in 193 \AA. Unlike N1 and N2, these two jets do not show any clear indication in the 131 \AA\ passband. The smaller size of these jets compared to N1 and N2 could be one of the reasons behind their feeble signatures in AIA passbands. These two jets also appear multi-thermal in nature with dominant contribution in the cooler channels of AIA. It should also be noted that N8 could not be distinctly identified in any AIA passband. This could be due to its relatively smaller size and short-lived nature (8 s), such that it could not be seen with the spatio-temporal resolution of AIA.}

\begin{figure*}
    \centering
    \includegraphics[width=1.05\textwidth,clip=]{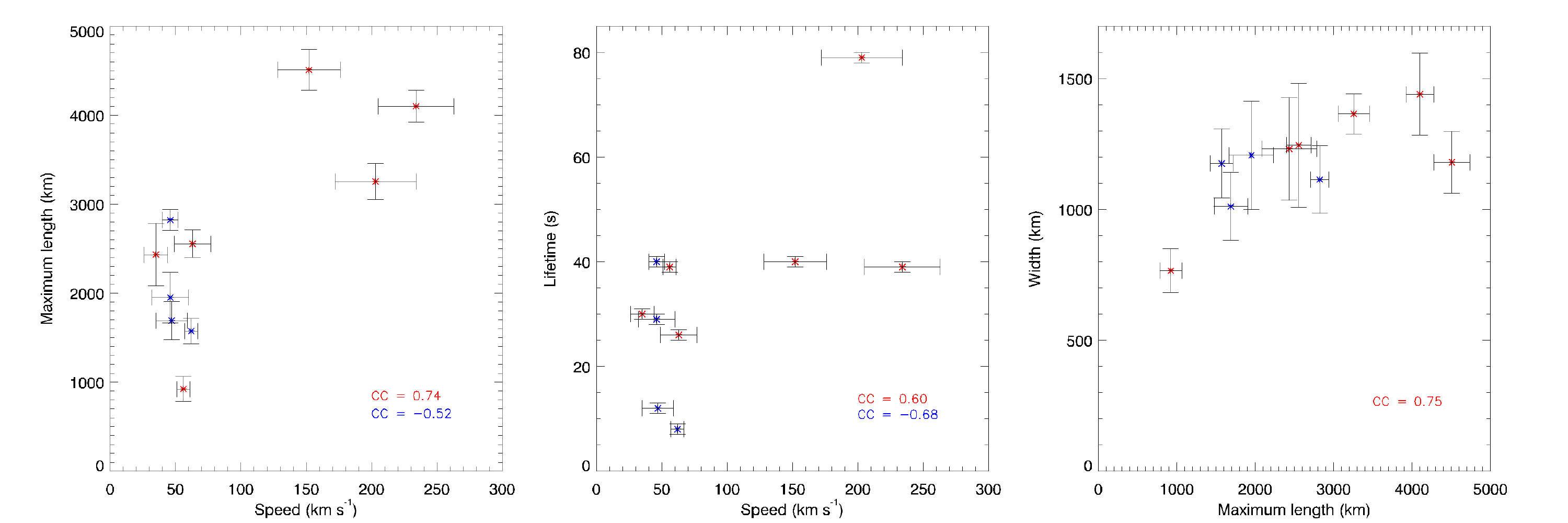}
    \caption{Relation between estimated parameters of reconnection nanojets where the outward jets are shown in red colour while the blue represents the inward moving ones. The correlation coefficients for the individual classes are shown in their respective colors. For the relation between length and width, negligible correlation is found for the inward directed jets.}
    \label{fig:stats}
\end{figure*}

{Time evolution images of other nanojets listed in Table \ref{tab:jet_prop} are included in Appendix \ref{sec: others}.}

\subsection{Correlation between physical parameters of nanojets}

{We studied the correlation of various estimated physical parameters of the identified nanojets. The left and middle panels of Figure \ref{fig:stats} shows the relation of the jets' speed with the maximum length attained by them and their lifetimes, respectively. The right panel of the same Figure shows the relation between the dimensions (length and width) of the jets. The outward and inward jets are shown in red and blue colour respectively. The error bars for the speed, length, and width corresponds to the 1$\sigma$ value mentioned in Table \ref{tab:jet_prop}, while that for lifetime is the temporal resolution of the Hi-C 2.1 instrument (4 s).}

{From the plots in left and middle panel it can be seen that these parameters are positively correlated for the outward jets, but negatively correlated for inward jets. We found a similar negative correlation of 45\% and 49\% was found for speed-length and speed-lifetime for the inward jets listed by \citet{2022ApJ...934..190S}. It can be seen in the right panel of the Figure \ref{fig:stats} that there is a good correlation of $\approx$75\% between the length and widths of the outward jets. However, for the observed sample of inward jets, a negligible correlation was found between the two parameters. A similar, however unreported, strong correlation of $\approx$82\% was found for such jets listed by \citet{2022ApJ...934..190S}.}

\section{Conclusions and Discussion} 
    \label{sec:conclusion}
    
We identify and report reconnection nanojets from the coronal loop associated with AR 12712 observed during the second flight of Hi-C in 172 \AA\ passband with high spatio-temporal resolution. {We identified and studied ten nanojets four of which had direction inward while six were directed outwards with respect to the direction of the radius of the curvature of the coronal loop. These jets were observed with locations at different regions in the coronal loops unlike in clusters as reported by \citet{2022ApJ...934..190S}. Adding to the previous studies of \citet{Antolin2021NatAs,2022ApJ...934..190S} where only two outward directed nanojets were found, we identified and report a few more of these nanojets in addition to the inward population.} The presence of outward jets with assymetric nature were predicted by the numerical MHD models but with a less probability as compared to the inward moving ones \citep{Pagano2021A&A}. 

To confirm that these jets originate from the loop and not from the photosphere, we also used HMI magnetograms and estimated an average magnetic field at the location of the base of the jets. We found that the location appears to be mostly quiet region with average magnetic flux at the noise level below 17 G \citep{Pesnell2012SoPh}.

{For the observed sample of Hi-C 2.1 nanojets we estimated the maximum length, and width of the spire corresponding to these jets, along with speed and their lifetime, and identified the correlations between various parameters. We found that the length and width varies from $\approx$900 km to $\approx$4500 km and $\approx$700 km to $\approx$1500 km respectively for outward directed jets while the inward jets have length in the range from $\approx$1500 km to $\approx$2800 km with widths $\approx$1100 km. These jets have dimensions more than the ones observed by \citet{Antolin2021NatAs} and lie at the tail of the distribution presented in \citet{2022ApJ...934..190S}.

For the outward jets we estimated the speeds from as low as 35 km s$^{-1}$ to more than 200 km s$^{-1}$. The speed for the inward jets ranges 40-60 km s$^{-1}$ implying their sub-sonic nature. A majority of the nanojets observed in previous studies had speeds in the range of few 100-200 km s$^{-1}$ with few jets listed by \citet{2022ApJ...934..190S} having speed below 100 km s$^{-1}$. It is worth noting that three of the outward moving jets have speed more than 150 km s$^{-1}$ which is comparable to the speed of inward directed jets in the earlier observational studies. This in contrary to the previous results that explained the outward jets are expected to be shorter and slower than the inward jets \citet{Antolin2021NatAs}.}

{The lifetime of the identified jets ranges shortest from 9 s to the longest duration of 79 s. { Average lifetime of outward jets is $\approx$42 s and the inward jets is $\approx$24 s}. The {outward} jets are long-lived features as compared to the $\le$15 s and $\approx$25 s lifetime jets reported in  \citet{Antolin2021NatAs,2022ApJ...934..190S}.} These jets are of short duration and less than half the length of jets reported in another study based on Hi-C observations by \citet{Panesar2019ApJ} where the jets had their origin due to photospheric magnetic flux cancellation. These nanojets are of about less than half the length but are of similar widths when compared with microjets observations by \citet{HouMicrojet2021ApJ} and jet-like campfires and EUV dots reported by \citet{Panesar2021ApJ} and \citet{Tiwari2022ApJ} respectively using the Solar Orbiter EUI data. {\bf The lifetime of the observed nanojets is on average smaller than the small-scale jet-like features in aforementioned studies.}
This shows that a wide spectrum of the jets exists in the solar atmosphere. A few of the nanojets are also observed in multiple passbands of AIA in 304, 171, 131, and 193 \AA. This suggests that the jets are multithermal in nature with temperature varying from those of upper transition region to hotter corona. 
% $\sim$40 s which is approximately two times more than that of inward moving jets observed earlier ($\le$15 s). On the other hand the plane of sky speed of these outward jets are measured to be 154 and 234 km s$^{-1}$ comparable to the inward ones reported in \citet{Antolin2021NatAs}. These jets are of short duration and less than half the length of jets reported in another study based on Hi-C observations by \citet{Panesar2019ApJ} where the jets had their origin due to photospheric magnetic flux cancellation. This shows that a wide spectrum of the jets exists in the solar atmosphere. 

{We also found that certain parameters associated with these jets are well correlated. Based on a sample of ten jets presented in this study, it turned out that faster outward jets have longer spire lengths and lasts longer whereas the inward faster jets are shorter in length as well as duration. A similar trend was also observed when these parameters of jets from \citet{2022ApJ...934..190S} were correlated for comparison. The outward jets also showed a good agreement for their physical dimensions which was not obtained for the inward jets. However, when these quantities from a larger sample of previous study was considered, a similar relation was also seen for the inward jets which was unreported earlier.}

% Upon performing a DEM analysis, it was found that the coronal temperatures at the location of these jets are of the order of 10$^{6.3}$ K suggesting that either these jets have plasma of coronal temperature. \citet{Antolin2021NatAs} reported a similar temperature range upon DEM for the observed inward moving jets.
% The rate of occurrence of outward nanojets for the 5 minutes of Hi-C observation is 0.4 per minute which is an order less than the inward ones reported by \citet{Antolin2021NatAs} as $\sim$15 per minute. 
This study suggests even though the preferential direction of jets in the loops may be inwards but the outward moving jets may not be so uncommon in the solar atmosphere. {This study could be extended to look for more reconnection nanojets using the data from EUI High Resolution Imager (HRI) of the Solar Orbiter.} Investigation of high spatio-temporal studies from IRIS and future missions of like Hi-C may help build a statistics of such reconnection jets allowing us to understand their role in heating the solar atmosphere at small scales. 
Such studies could be further complemented with the high-resolution imaging and spectroscopy based future missions such as  Multi-slit Solar Explorer \citep[MUSE;][]{MUSE2020ApJ, DePontieu2021arXiv, Cheung2021arXiv}, Solar-C EUV High-Throughput Spectroscopic Telescope \citep[EUVST;][]{EUVST2019SPIE} covering a wide range of temperatures from transition region to corona.

\begin{acknowledgments}
{We thank the anonymous referee for their valuable comments and suggestions.}
We acknowledge NASA/SDO team to make AIA and HMI data for open access. SDO is a mission for NASA's Living With a Star (LWS) program. We acknowledge the High-resolution Coronal Imager (Hi-C 2.1) instrument team for making the second re-flight data available under NASA proposal 17-HTIDS17\_2-003. MSFC/NASA led the mission with partners including the Smithsonian Astrophysical Observatory, the University of Central Lancashire, and the Lockheed Martin Solar and Astrophysics Laboratory. Hi-C 2.1 was launched out of the White Sands Missile Range on 2018 May 29.
\end{acknowledgments}

%% This command is needed to show the entire author+affiliation list when
%% the collaboration and author truncation commands are used.  It has to
%% go at the end of the manuscript.
%\allauthors

%% Include this line if you are using the \added, \replaced, \deleted
%% commands to see a summary list of all changes at the end of the article.
%\listofchanges

\appendix

\section{Hi-C Images of Other Nanojets listed in Table 1}
\label{sec: others}

\begin{figure*}[!ht]
    \centering
    \includegraphics[width=1\textwidth,clip=]{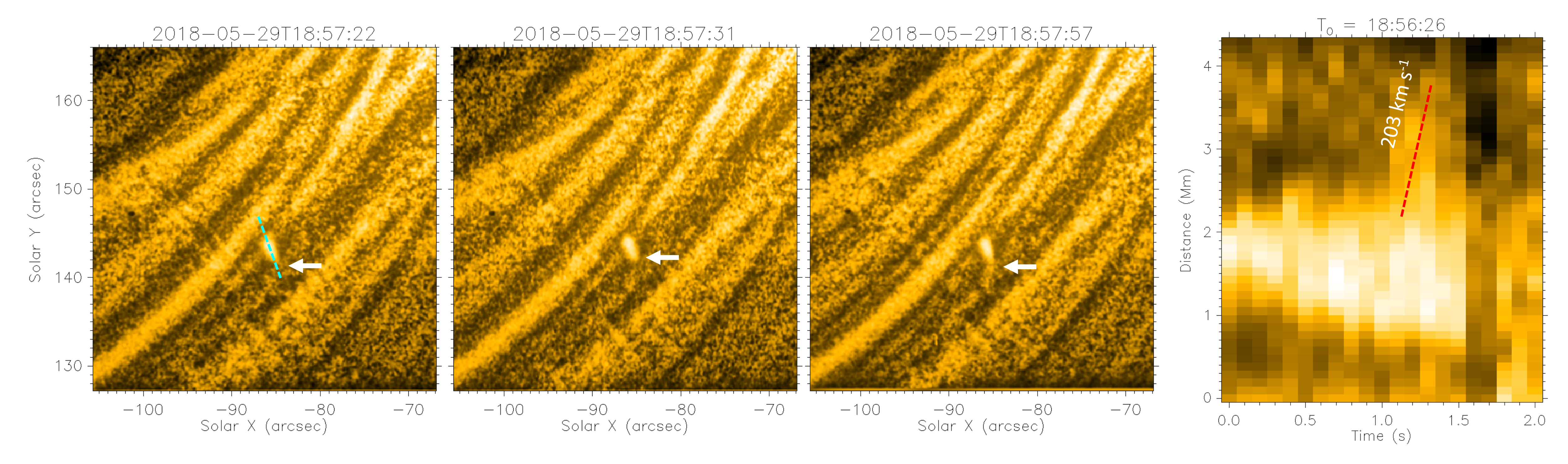}
    \caption{Time evolution an outward moving nanojet, N3. The first three panels show snapshots for nanojet 3. The white arrow points to the location of the jet. The time-distance plot shown in the last panel is generated along the dashed cyan line shown in {\it left} panel. The starting time shown as T$_0$ at the top of the plot. \\ ({Animation of this Figure is available showing the time evolution of the jet.)}}
    \label{fig:n3}
\end{figure*}

\begin{figure*}[!h]
    \centering
    \includegraphics[width=1\textwidth,clip=]{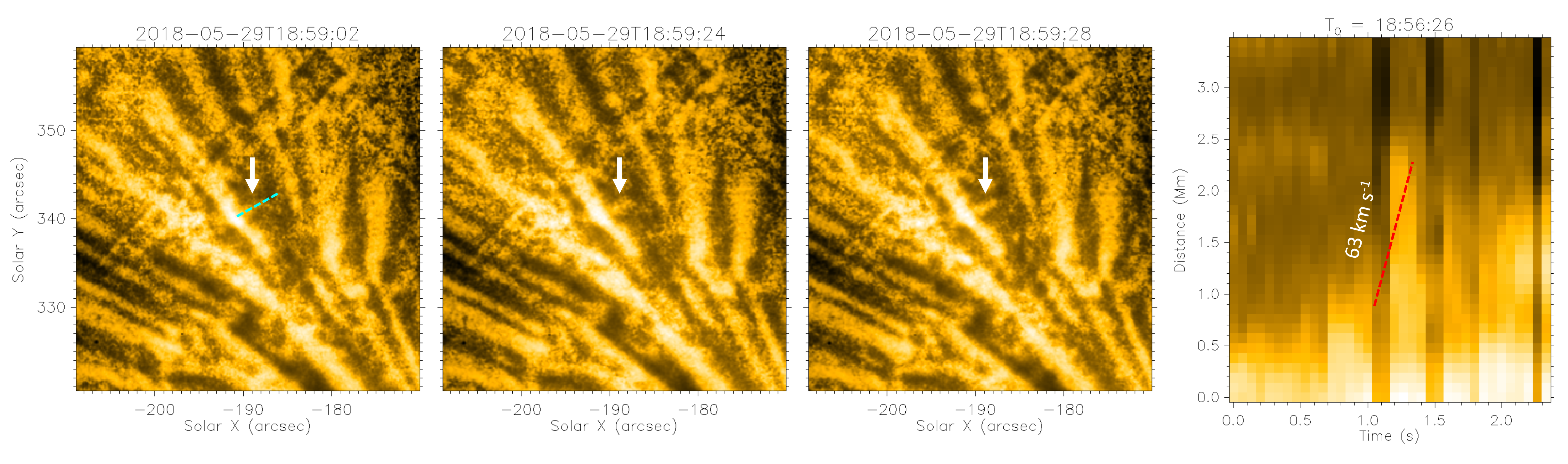}
    \caption{Time evolution an outward moving nanojet, N5. The first three panels show snapshots for nanojet 5. The white arrow points to the location of the jet. The time-distance plot shown in the last panel is generated along the dashed cyan line shown in {\it left} panel. The starting time shown as T$_0$ at the top of the plot. \\ ({Animation of this Figure is available showing the time evolution of the jet.)}}
    \label{fig:n3}
\end{figure*}

\begin{figure*}[!h]
    \centering
    \includegraphics[width=1\textwidth,clip=]{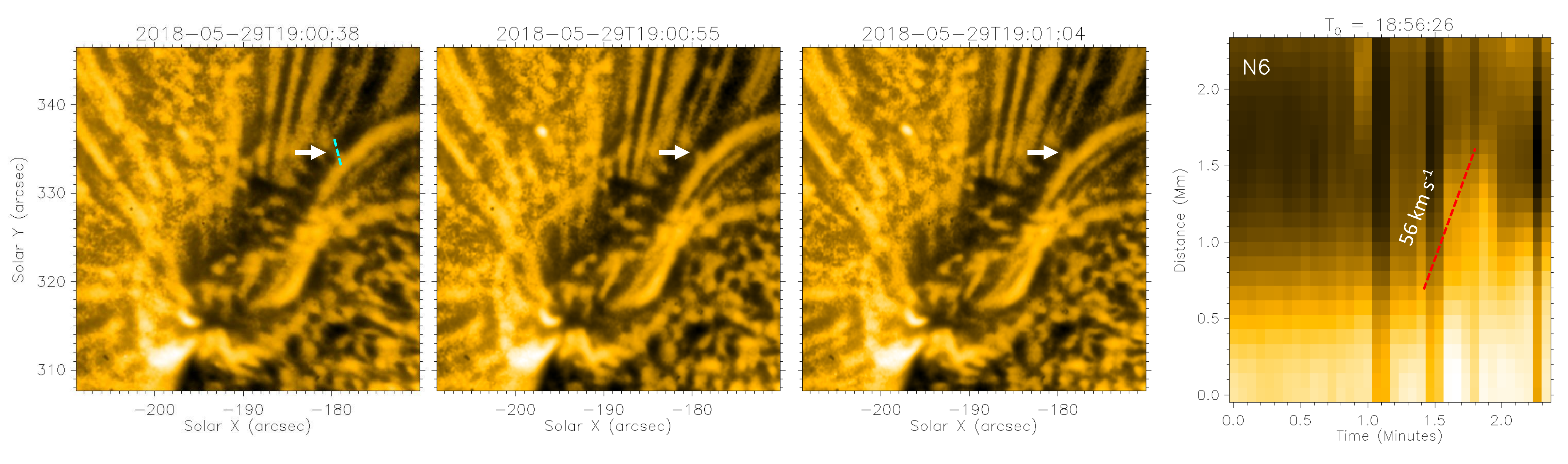}
    \caption{Time evolution an outward moving nanojet, N6. The first three panels show snapshots for nanojet 6. The white arrow points to the location of the jet. The time-distance plot shown in the last panel is generated along the dashed cyan line shown in {\it left} panel. The starting time shown as T$_0$ at the top of the plot. \\ ({Animation of this Figure is available showing the time evolution of jet.)}}
    \label{fig:n3}
\end{figure*}

\begin{figure*}[!h]
    \centering
    \includegraphics[width=1\textwidth,clip=]{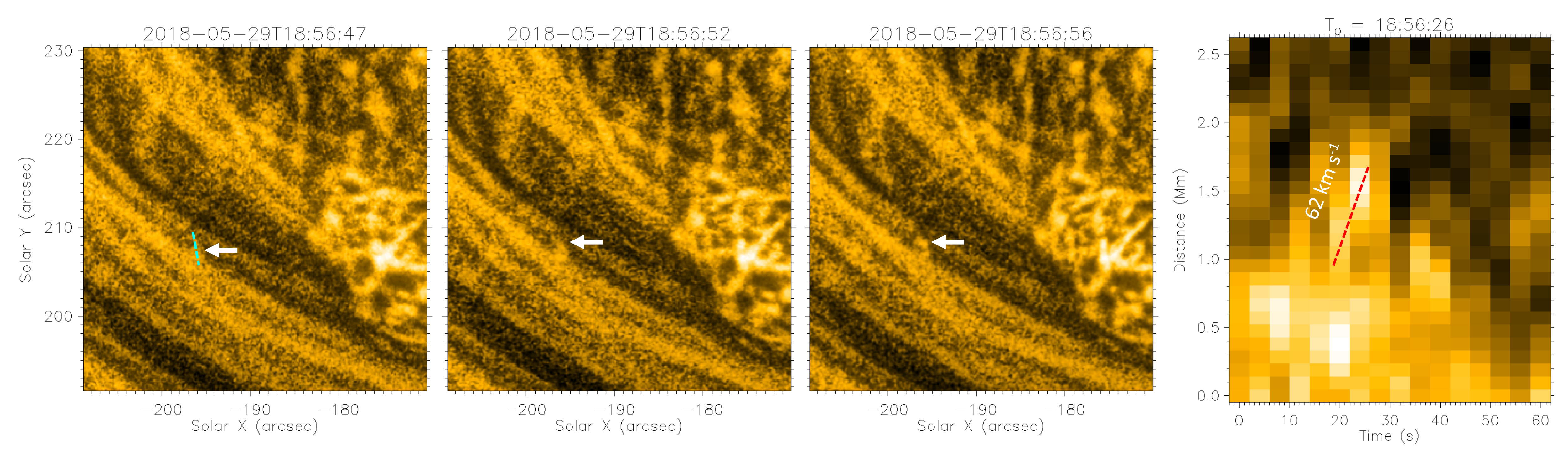}
    \caption{Time evolution an inward moving nanojet, N9. The first three panels show snapshots for nanojet 9. The white arrow points to the location of the jet. The time-distance plot shown in the last panel is generated along the dashed cyan line shown in {\it left} panel. The starting time shown as T$_0$ at the top of the plot. \\ ({Animation of this Figure is available showing the time evolution of the jet.)}}
    \label{fig:n3}
\end{figure*}

\begin{figure*}[!h]
    \centering
    \includegraphics[width=1\textwidth,clip=]{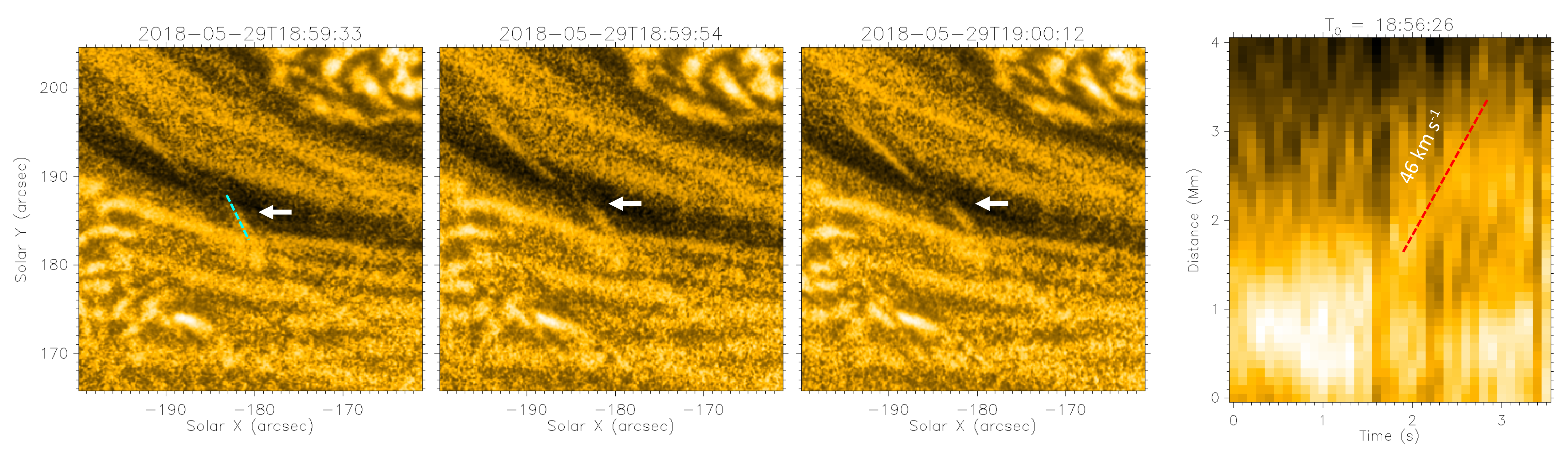}
    \caption{Time evolution an inward moving nanojet, N10. The first three panels show snapshots for nanojet 3. The white arrow points to the location of the jet. The time-distance plot shown in the last panel is generated along the dashed cyan line shown in {\it left} panel. The starting time shown as T$_0$ at the top of the plot. \\ ({Animation of this Figure is available showing the time evolution of the jet.)}}
    \label{fig:n3}
\end{figure*}

\bibliography{bibli}{}
\bibliographystyle{aasjournal}

\end{document}